\documentclass[11pt,a4paper,reqno]{amsart}

\usepackage[margin=0.8in,footskip=0.25in]{geometry}
\usepackage{amsmath, amsthm, amssymb}
\usepackage{graphicx}

\usepackage{pslatex}
\usepackage{array}
\usepackage[square,numbers]{natbib}
\usepackage{amsmath,systeme}
\usepackage{hyperref}

\date{}

\begin{document}

\title[Modelling the dynamics of cross-border ideological competition]
{Modelling the dynamics of cross-border ideological competition}

\author{Jose Segovia-Martin$^1$}
\address{$^1$ Complex Systems Institute of Paris Ile-de-France (ISC-PIF)}
\address{$^2$ Centre national de la recherche scientifique (CNRS)}
\address{$^3$ School of Collective Intelligence (M6 Polytechnic University)}
\email{Jose.Segovia@um6p.ma}

\subjclass[2010]{37B55, 34A34.}

\keywords{Dynamical Systems $|$ Computational social science $|$ Social diffusion $|$ Complex contagion $|$ Influence $|$ Ideology $|$ Political science}

\begin{abstract}
Individuals are increasingly exposed to news and opinion from beyond national borders in a world that is becoming more and more globalised. This news and opinion is often concentrated in clusters of ideological homophily such as political parties, factions or interest groups. But how does exposure to cross-border information affect the diffusion of ideas across national and ideological borders?
Here we develop a non-linear mathematical model for the cross-border spread of two ideologies by using an epidemiological approach. The populations of each country are assumed to be a constant and homogeneously mixed. We solve the system of differential equations numerically and show how small changes in the influence of a minority ideology can trigger shifts in the global political equilibrium.
\end{abstract}

\maketitle

\section{Main}

Existing models of ideology dynamics have focused on the transmission and evolution of political ideas within a single voting population \cite{misra2012simple, calderon2005epidemiological, fieldhouse2007strategic, segovia2021synchronising, nyabadza2016modelling, petersen1991stability,khan2000hopf}. In contrast, the model we describe here idealises ideologies as fixed and as competing with each other for supporters both within and across borders. We also assume constant and homogeneous mixed populations with no spatial or social structure. Agents in our model can only support one ideology (party or political tendency) at any given moment in time.

Let us consider two countries whose respective homogeneously distributed populations are $N_1$ and $N_2$. The population of country 1 $N_1$ consists of three sets of agents, namely: (i) without ideological or political affiliation $V_1$, (ii) with ideology or political affiliation $B$, (iii) with ideology or political affiliation $C$. Similarily, the population of country 2 $N_2$ consists of three sets of agents, namely (i) without ideological or political affiliation $V_2$, (ii) with ideology or political affiliation $D$, (iii) with ideology or political affiliation $E$. Now, assume that $B$ and $D$ have the same ideology, and that $C$ and $E$ also share the same ideology. These two groups form two blocks of competing ideologies (e.g pro and anti-tax, pro and anti-vaccination, pro and anti-immigration, etc...). 

In our model of cross-border influence we will allow that each of the blocks is able to recruit supporters (voters) both within and outside its borders, so that for example, $B$ will be able to recruit supporters from $V_1$ and $C$ within its borders, but will also be able to exert influence outside its borders by recruiting agents from $V_2$ and $E$ towards $D$. Similarly, $C$, $D$, $E$ will be able to recruit supporters for themselves and for their ideological partners beyond their borders.

On the other hand, consider there are rates $\mu_2$ and $\mu_4$ at which agents $\mu_2 V_1$ and $\mu_4 V_2$ cease to be potential voters of countries 1 and 2 respectively due to deth or migration. Similarly, we have death ratios $\mu_B$, $\mu_C$, $\mu_D$ and $\mu_E$ for each of the political groups, all labelled with their respective sub-indexes. A system with no gains or losses of citizens over time will keep $\mu$ constant and of equal value across all population groups.

Now, consider that there is a rate $\mu_1$ at which agents $\mu_1 N$ enter the system in country 1, and similarly a rate $\mu_2$ at which agents $\mu_2 N$ enter the system in country 2. This parameter can be thought of as the rate at which individuals reach the legal age or at which they attain the necessary civic knowledge, skills, cognitive ability and right to vote.

Parameter $k_1$ stands for the average number of contacts of agents of $B$ with agents of $V_1$ and $p_1$ is the probability of $B$ convincing another agent per contact. This means that the term $k_1 p_1 V_1 \frac{B}{N_1}$ stands for the rate of agents that move from $V_1$ to $B$, where $\frac{B}{N_1}$ is the chance of coming into contact with the members of $B$ in country 1 (i.e. the relative weight of B in the population). Similarly, the term $k_2 p_2 V_1 \frac{C}{N_1}$ is the rate of agents that move from $V_1$ to $C$, the term $k_3 p_3 V_2 \frac{D}{N_2}$ is the rate of agents that move from $V_2$ to $D$ and $k_4 p_4 V_2 \frac{E}{N_2}$ is the rate of agents that move from $V_2$ to $E$. Also, agents of $V_1$ may decide to join $B$, not because of $B$ 's direct influence, but as a consequence of a foreign influence of the same nature as $B$, in our case $D$. The transfer of agents from $V_1$ to $B$ due to the influence of $D$ occurs at rate $(1-k_1 p_1) k_3 p_3 V_1 \frac{D}{N_2}$. Likewise, $E$ can capture agents from $V_1$ to $C$ at rate $(1-k_2 p_2) k_4 p_4 V_1 \frac{E}{N_2}$. In country 2, we have the same mechanism reversed for $D$ and $E$.

But ideological or political affiliation can fade over time. In the model this loss can be described by a leakage of agents from the $B$ category back to $V_1$ at rate $\gamma_1 B$, and from $C$ to $V_1$ at rate $\gamma_2 C$. In country 2, we have the same, there is a leakage of agents from $D$ to $V_2$ at rate $\gamma_3 D$ and from $E$ to $V_2$ at rate $\gamma_4 E$.

Finally, let $\phi_1$ and $\phi_2$ be the per capita recruitment capacity of $B$ from $C$ and of $C$ from $B$ respectively. Similarily, $\phi_3$ and $\phi_4$ are the per capita recruitment capacity of $D$ from $E$ and of $E$ from $D$ respectively. Therefore, agents of $C$ decide to go to $B$ due to $B$'s influence at rate $\phi_1 C \frac{B}{N_1}$ and due to $D$'s influence at rate $(1-\phi_1) \phi_3 C \frac{D}{N_2}$, while agents of $B$ decide to go to $C$ due to $C$'s influence at rate $\phi_2 B \frac{C}{N_1}$ and due to $E$'s influence at rate $(1-\phi_2) \phi_4 B \frac{E}{N_2}$. Likewise, in country 2 agents of $E$ move to $D$ due to $D$'s influence at rate $\phi_3 E \frac{D}{N_2}$ and due to $B$'s influence at rate $(1-\phi_3) \phi_1 E \frac{B}{N_1}$, while agents of $D$ decide to go to $E$ due to $E$'s influence at rate $\phi_4 D \frac{E}{N_2}$ and due to $C$'s influence at rate $(1-\phi_4) \phi_2 D \frac{C}{N_1}$.

In accordance with the parameters, terms and assumptions described above, the governing differential equations of the model can be written as follows \footnote[1]{$N_1, V_1, B, C, N_2, V_2, D, E$ are all dependent on time. For brevity of notation, the time dependencies of $N_1(t), V_1(t), B(t), C(t), N_2(t), V_2(t), D(t), E(t)$ are not made explicit in the equations throughout the paper.}:
\begin{equation}
\systeme{
\small
\frac{dV_1}{dt}=\mu_1 N_1 - k_1 p_1 V_1 \frac{B}{N_1} - (1-k_1 p_1) k_3 p_3 V_1 \frac{D}{N_2} - k_2 p_2 V_1 \frac{C}{N_1} - (1-k_2 p_2) k_4 p_4 V_1 \frac{E}{N_2} - \mu_2 V_1 + \gamma_1 B + \gamma_2 C ,
\small
\frac{dB}{dt}= k_1 p_1 V_1 \frac{B}{N_1} + (1-k_1 p_1) k_3 p_3 V_1 \frac{D}{N_2} - \phi_2 B \frac{C}{N_1} - (1-\phi_2) \phi_4 B \frac{E}{N_2} + \phi_1 C \frac{B}{N_1} + (1-\phi_1) \phi_3 C \frac{D}{N_2} - \mu_B B - \gamma_1 B ,
\small
\frac{dC}{dt}= k_2 p_2 V_1 \frac{C}{N_1} + (1-k_2 p_2) k_4 p_4 V_1 \frac{E}{N_2} - \phi_1 C \frac{B}{N_1} - (1-\phi_1) \phi_3 C \frac{D}{N_2} + \phi_2 B \frac{C}{N_1} + (1-\phi_2) \phi_4 B \frac{E}{N_2} - \mu_C C - \gamma_2 C ,
\small
\frac{dV_2}{dt}=\mu_3 N_2 - k_3 p_3 V_2 \frac{D}{N_2} - (1-k_3 p_3) k_1 p_1 V_2 \frac{B}{N_1} - k_4 p_4 V_2 \frac{E}{N_2} - (1-k_4 p_4) k_2 p_2 V_2 \frac{C}{N_1} - \mu_4 V_2 + \gamma_3 D + \gamma_4 E ,
\small
\frac{dD}{dt}= k_3 p_3 V_2 \frac{D}{N_2} + (1-k_3 p_3) k_1 p_1 V_2 \frac{B}{N_1} - \phi_4 D \frac{E}{N_2} - (1-\phi_4) \phi_2 D \frac{C}{N_1} + \phi_3 E \frac{D}{N_2} + (1-\phi_3) \phi_1 E \frac{B}{N_1} - \mu_D D - \gamma_3 D ,
\small
\frac{dE}{dt}= k_4 p_4 V_2 \frac{E}{N_2} + (1-k_4 p_4) k_2 p_2 V_2 \frac{C}{N_1} - \phi_3 E \frac{D}{N_2} - (1-\phi_3) \phi_1 E \frac{B}{N_1} + \phi_4 D \frac{E}{N_2} + (1-\phi_4) \phi_2 D \frac{C}{N_1} - \mu_E E - \gamma_4 E }
\end{equation}

where $N_1=V_1 + B + C$ being $V_1 (0) > 0, B(0) \geq 0, C(0) \geq 0$ and $N_2=V_2 + D + E$ being $V_2 (0) > 0, D(0) \geq 0, E(0) \geq 0$. Adding the equations we see that $dN_1/dt=0$ and $dN_2/dt=0$.

Given that $k_1$ stands for the average number of contacts of members of $B$ with members of $V_1$ per unit time, and that $p_1$ is the probability of $B$ convincing another agent per contact, then we have that the per capita recruitment rate of $B$ from $V_1$ is $\beta_1 = p_1 k_1$. The same applies to the rest of the equations, where: $\beta_2 = p_2 k_2$, $\beta_3 = p_3 k_3$, $\beta_4 = p_4 k_4$.Therefore, the model can be reduced to the following system:

\begin{equation}
\systeme{
\small
\frac{dV_1}{dt}=\mu_1 N_1 - \beta_1 V_1 \frac{B}{N_1} - (1-\beta_1) \beta_3 V_1 \frac{D}{N_2} - \beta_2 V_1 \frac{C}{N_1} - (1-\beta_2) \beta_4 V_1 \frac{E}{N_2} - \mu_2 V_1 + \gamma_1 B + \gamma_2 C ,
\small
\frac{dB}{dt}= \beta_1 V_1 \frac{B}{N_1} + (1-\beta_1) \beta_3 V_1 \frac{D}{N_2} - \phi_2 B \frac{C}{N_1} - (1-\phi_2) \phi_4 B \frac{E}{N_2} + \phi_1 C \frac{B}{N_1} + (1-\phi_1) \phi_3 C \frac{D}{N_2} - \mu_B B - \gamma_1 B ,
\small
\frac{dC}{dt}= \beta_2 V_1 \frac{C}{N_1} + (1-\beta_2) \beta_4 V_1 \frac{E}{N_2} - \phi_1 C \frac{B}{N_1} - (1-\phi_1) \phi_3 C \frac{D}{N_2} + \phi_2 B \frac{C}{N_1} + (1-\phi_2) \phi_4 B \frac{E}{N_2} - \mu_C C - \gamma_2 C ,
\small
\frac{dV_2}{dt}=\mu_3 N_2 - \beta_3 V_2 \frac{D}{N_2} - (1-\beta_3) \beta_1 V_2 \frac{B}{N_1} - \beta_4 V_2 \frac{E}{N_2} - (1-\beta_4) \beta_2 V_2 \frac{C}{N_1} - \mu_4 V_2 + \gamma_3 D + \gamma_4 E ,
\small
\frac{dD}{dt}= \beta_3 V_2 \frac{D}{N_2} + (1-\beta_3) \beta_1 V_2 \frac{B}{N_1} - \phi_4 D \frac{E}{N_2} - (1-\phi_4) \phi_2 D \frac{C}{N_1} + \phi_3 E \frac{D}{N_2} + (1-\phi_3) \phi_1 E \frac{B}{N_1} - \mu_D D - \gamma_3 D ,
\small
\frac{dE}{dt}= \beta_4 V_2 \frac{E}{N_2} + (1-\beta_4) \beta_2 V_2 \frac{C}{N_1} - \phi_3 E \frac{D}{N_2} - (1-\phi_3) \phi_1 E \frac{B}{N_1} + \phi_4 D \frac{E}{N_2} + (1-\phi_4) \phi_2 D \frac{C}{N_1} - \mu_E E - \gamma_4 E }
\end{equation}

Now, because the transfer of agents between $B$ and $C$ due to their influence within the country results in a net amount of exchange, it follows that $\phi_2 - \phi_1 = \phi_{w1}$. The same for $D$ and $E$, where we have: $\phi_4 - \phi_3 = \phi_{w2}$. Following the same reasoning, we also observe that there is a net transfer of agents due to cross-border influence, therefore $(1-\phi_2) \phi_4 - (1-\phi_1) \phi_3 = \phi_{b1}$ and $(1-\phi_4) \phi_2 - (1-\phi_3) \phi_1 = \phi_{b2}$. After this reduction, our system can be written as follows:

\begin{equation}
\systeme{
\small
\frac{dV_1}{dt}=\mu_1 N_1 - \beta_1 V_1 \frac{B}{N_1} - (1-\beta_1) \beta_3 V_1 \frac{D}{N_2} - \beta_2 V_1 \frac{C}{N_1} - (1-\beta_2) \beta_4 V_1 \frac{E}{N_2} - \mu_2 V_1 + \gamma_1 B + \gamma_2 C ,
\small
\frac{dB}{dt}= \beta_1 V_1 \frac{B}{N_1} + (1-\beta_1) \beta_3 V_1 \frac{D}{N_2} - \phi_{w1} B \frac{C}{N_1} - \phi_{b1} B \frac{E}{N_2} - \mu_B B - \gamma_1 B ,
\small
\frac{dC}{dt}= \beta_2 V_1 \frac{C}{N_1} + (1-\beta_2) \beta_4 V_1 \frac{E}{N_2} + \phi_{w1} B \frac{C}{N_1} + \phi_{b1} B \frac{E}{N_2} - \mu_C C - \gamma_2 C ,
\small
\frac{dV_2}{dt}=\mu_3 N_2 - \beta_3 V_2 \frac{D}{N_2} - (1-\beta_3) \beta_1 V_2 \frac{B}{N_1} - \beta_4 V_2 \frac{E}{N_2} - (1-\beta_4) \beta_2 V_2 \frac{C}{N_1} - \mu_4 V_2 + \gamma_3 D + \gamma_4 E ,
\small
\frac{dD}{dt}= \beta_3 V_2 \frac{D}{N_2} + (1-\beta_3) \beta_1 V_2 \frac{B}{N_1} - \phi_{w2} D \frac{E}{N_2} - \phi_{b2} D \frac{C}{N_1} - \mu_D D - \gamma_3 D ,
\small
\frac{dE}{dt}= \beta_4 V_2 \frac{E}{N_2} + (1-\beta_4) \beta_2 V_2 \frac{C}{N_1} + \phi_{w2} D \frac{E}{N_2} + \phi_{b2} D \frac{C}{N_1} - \mu_E E - \gamma_4 E }
\end{equation}

And so after division we obtain the following differential equations:

\begin{equation}
\systeme{
\small
\frac{dv_1}{dt}=\mu_1 - \beta_1 v_1 b - (1-\beta_1) \beta_3 v_1 d - \beta_2 v_1 c - (1-\beta_2) \beta_4 v_1 e - \mu_2 v_1 + \gamma_1 b + \gamma_2 c ,
\small
\frac{db}{dt}= \beta_1 v_1 b + (1-\beta_1) \beta_3 v_1 d - \phi_{w1} b c - \phi_{b1} b e - \mu_B b - \gamma_1 b ,
\small
\frac{dc}{dt}= \beta_2 v_1 c + (1-\beta_2) \beta_4 v_1 e + \phi_{w1} b c + \phi_{b1} b e - \mu_C c - \gamma_2 c ,
\small
\frac{dv_2}{dt}=\mu_3 - \beta_3 v_2 d - (1-\beta_3) \beta_1 v_2 b - \beta_4 v_2 e - (1-\beta_4) \beta_2 v_2 c - \mu_4 v_2 + \gamma_3 d + \gamma_4 e ,
\small
\frac{dd}{dt}= \beta_3 v_2 d + (1-\beta_3) \beta_1 v_2 b - \phi_{w2} d e - \phi_{b2} d c - \mu_D d - \gamma_3 d ,
\small
\frac{de}{dt}= \beta_4 v_2 e + (1-\beta_4) \beta_2 v_2 c + \phi_{w2} d e + \phi_{b2} d c - \mu_E e - \gamma_4 e }
\end{equation}

Now, let us denote the equilibrium of the above system as ($v_1^*$, $b^*$, $c^*$, $v_2^*$, $d^*$, $e^*$) and therefore $v_1^*$ = $V_1^*$/$N_1$, $b^*$ = $B^*$/$N_1$, $c^*$ = $C^*$/$N_1$, $v_2^*$=$V_2^*$/$N_2$, $d^*$=$D^*$/$N_2$, $e^*$=$E^*$/$N_2$, where ($V_1^*$, $B^*$, $C^*$, $V_2^*$, $D^*$, $E^*$) represents the equilibrium of the unreduced system. Since the population of agents in each country remains constant as given by $N_1=V_1+B+C$ and $N_2=V_2+D+E$, we deduce $v_1 + b + c = 1$ and $v_2 + d + e = 1$ for the reduced system. Using this fact, the reduced model system will be given by the following four differential equations:

\begin{equation}
\systeme{
\small
\frac{db}{dt}= \beta_1 (1-b-c) b + (1-\beta_1) \beta_3 (1-b-c) d - \phi_{w1} b c - \phi_{b1} b e - \mu_B b - \gamma_1 b ,
\small
\frac{dc}{dt}= \beta_2 (1-b-c) c + (1-\beta_2) \beta_4 (1-b-c) e + \phi_{w1} b c + \phi_{b1} b e - \mu_C c - \gamma_2 c ,
\small
\frac{dd}{dt}= \beta_3 (1-d-e) d + (1-\beta_3) \beta_1 (1-d-e) b - \phi_{w2} d e - \phi_{b2} d c - \mu_D d - \gamma_3 d ,
\small
\frac{de}{dt}= \beta_4 (1-d-e) e + (1-\beta_4) \beta_2 (1-d-e) c + \phi_{w2} d e + \phi_{b2} d c - \mu_E e - \gamma_4 e }
\end{equation}

We conducted numerical simulations assuming constant $\mu$ = 0.017 and constant $\gamma$=0.01. We assume that the average individual acquires the right to vote at the age of 18 and spends about 60 years of his or her life with an active political life, i.e. $\mu$=1/60. We systematically manipulated parameters $k$ (average number of contacts per time per capita of a given party), parameters $p$ (capacity to convince of a given party) and $\phi$ (per capita recruitment of a given party). Following the logic of previous models, we keep recruitment capacity parameters $\beta$ and $\phi$ at realistically low values (0 to 0.1). For example, 1/25 = 0.04 means that 25 members of a political party are able to recruit 1 voter in a year.

\begin{figure*}[ht!]
\includegraphics[width=1\textwidth]{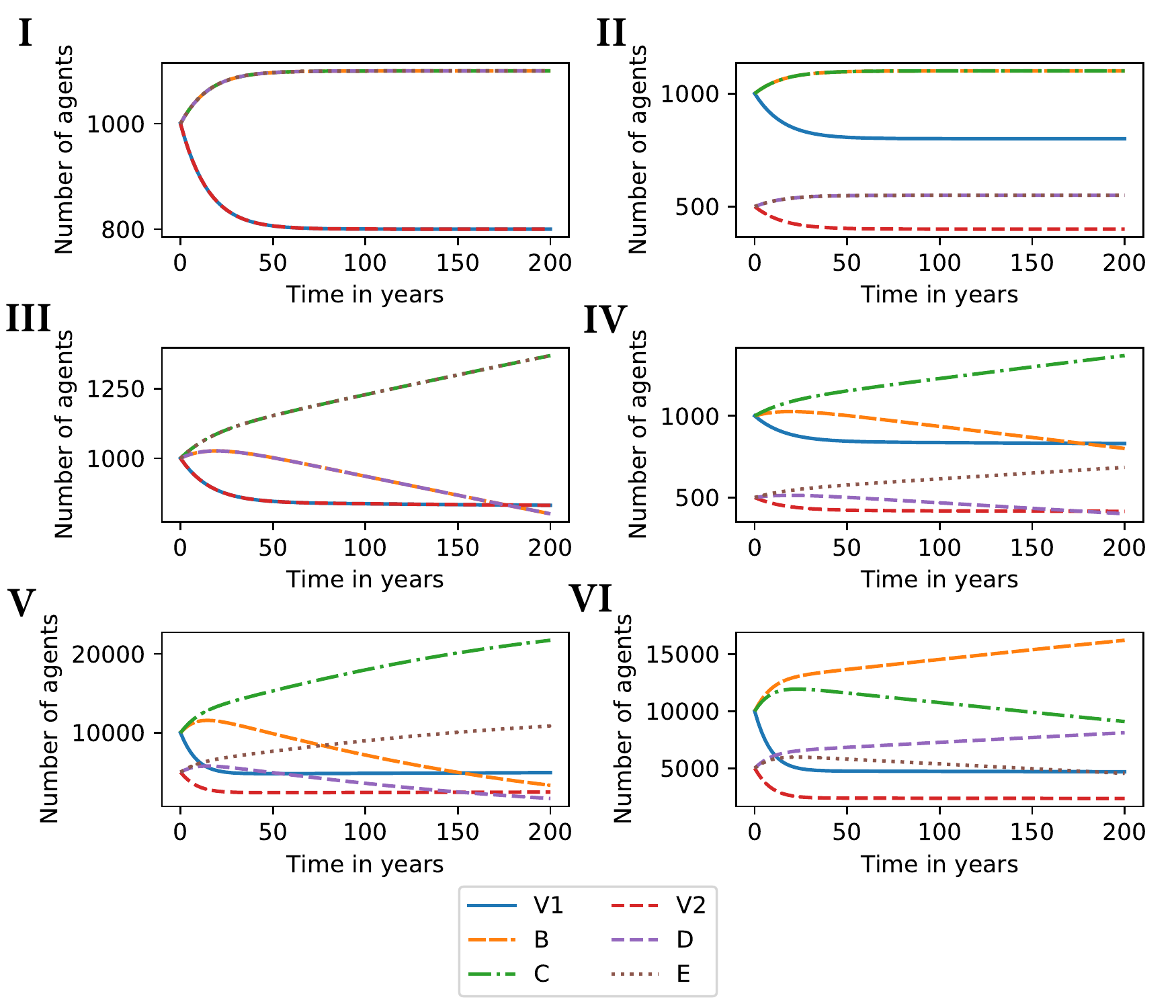}
  \centering
  \caption{Number of agents supporting each ideological bloc over time. I: Simulations for same initial population size ($V_1=B=C=V_2=D=E$) and same influence, with parameters $k=0.5$, $p=0.1$, $\phi=0.02$. II: Simulations for different initial population size ($V_1/2=B/2=C/2=V_2=D=E$) and same influence, with parameters $k=0.5$, $p=0.1$, $\phi=0.02$. III: Simulations for same initial population size ($V_1=B=C=V_2=D=E$) and different influence, with $k_1=0.4$ and parameters $k=0.5$, $p=0.1$, $\phi=0.02$. IV: Simulations for different initial population size ($V_1/2=B/2=C/2=V_2=D=E$) and different influence, with $k_1=0.4$ and parameters $k=0.5$, $p=0.1$, $\phi=0.02$. V and VI: Simulations for different initial population size ($V_1/2=B/2=C/2=V_2=D=E$) and different influence, with $k_1= 0.6$, $k_2= 0.4$, $k_3=0.6$, $k_4=0.6$, $p_1= 0.2$, $p_2= 0.1$, $p_3=0.1$, $p_4=0.2$, $\phi_1= 0.01$, $\phi_2= 0.03$, $\phi_4=0.01$. With these parameters, there is an inflection point around $\phi_3$=0.0267 whose values above and below determine the domains of the success function of one or the other competing ideology.}
  \label{fig:Figure_1}
  \end{figure*}

Our simulations show various equilibria, some with co-dominance of ideologies (Fig. \ref{fig:Figure_1} I \& II), and others with dominance of one ideology (Fig. \ref{fig:Figure_1} III \& IV). Interestingly, in some areas of the parameter space, a small change in the recruiting capacity of one of the parties can produce a total reversal of the balance of power. This butterfly effect is illustrated by the comparison of scenarios V and VI in Fig. \ref{fig:Figure_1}, which makes it clear that the system suffers from deterministic chaos. An increase as small as 0.005 in the initial recruitment capacity of the minority party $D$ from $E$ in country $N_2$ triggers a global ideological shift, revealing a high sensitivity of the system to initial conditions. According to our model, these seemingly imperceptible shifts in influence around the tipping point can consummate short-term political changes (within a few decades) and extinctions of once-dominant ideologies within a few hundred years.

Our model simulates a closed system with two hypothetical countries where political parties can exert influence within and across borders. In the real world this idealised situation does not exist. However, our model can inform how ideologies compete across national boundaries in an increasingly globalised world. For example, pro- and anti-democratic attitudes or pro- and anti-authoritarian views seem to behave as competing ideological blocks beyond national borders \cite{martins2020rise}. And today, more than ever before, the internet and social media have intensified this cross-border competition of ideologies on a global level. 

One take-home message that emerges from our model is that small changes in the influence of an ideology, a party or a minority opinion can trigger substantial political change in the medium to long term. Think, as an analogy, of the historical struggles for women's or ethnic minority rights and how many once minority ideas of equality and freedom have gradually percolated through society. But consider also the ease with which almost extinct totalitarian ideas are reborn and spread beyond national borders at certain historical moments.

So what can we learn from our model of political influence? The example we have illustrated here shows that small acts promoting a minority idea can trigger aggregate processes that eventually culminate in ideological change at the global level.

Our model works with spatially unstructured populations, but we know from previous studies that homogeneously and heterogeneously mixed populations can have different effects on the transmission of social information \cite{keeling2005networks, rahmandad2008heterogeneity, centola2015spontaneous, segovia2020network, walker2021maintenance}, so one avenue of future research will be to investigate the effect of different network structures on our model.

A more detailed analysis of the equilibrium and stability of the system will allow us to study the evolution of the system more precisely for relevant political and social scenarios.

\section*{Data Availability}

Electronic supplementary material and simulation code are available at: \\ https://github.com/jsegoviamartin/cross-border-ideological-competition

\section*{Acknowlegements}
I thank my colleagues from the School of Collective Intelligence (SCI) and the Complex Systems Institute of Paris (ISCPIF) for helpful discussions.

% Bibliography
%\section*{References}
\bibliographystyle{unsrtnat}
\bibliography{main}

\end{document}